\documentclass[10pt, conference]{IEEEtran}
\usepackage{cite}
\ifCLASSINFOpdf
\usepackage[pdftex]{graphicx}
\else
\fi
\usepackage{url}
\usepackage{flushend}
\usepackage[utf8]{inputenc}
\usepackage{german}
\usepackage[pdftex]{graphicx}
\usepackage{wrapfig}
\usepackage{color}
\usepackage{listings}
\usepackage{amsmath}
\usepackage{mathcomp}
\usepackage{array}
\usepackage[font=footnotesize,justification=raggedright,singlelinecheck=false]{caption}

\usepackage{caption}

\begin{document}

\newenvironment{itemize_positive}{%
\renewcommand{\labelitemi}{\includegraphics[height=8pt]{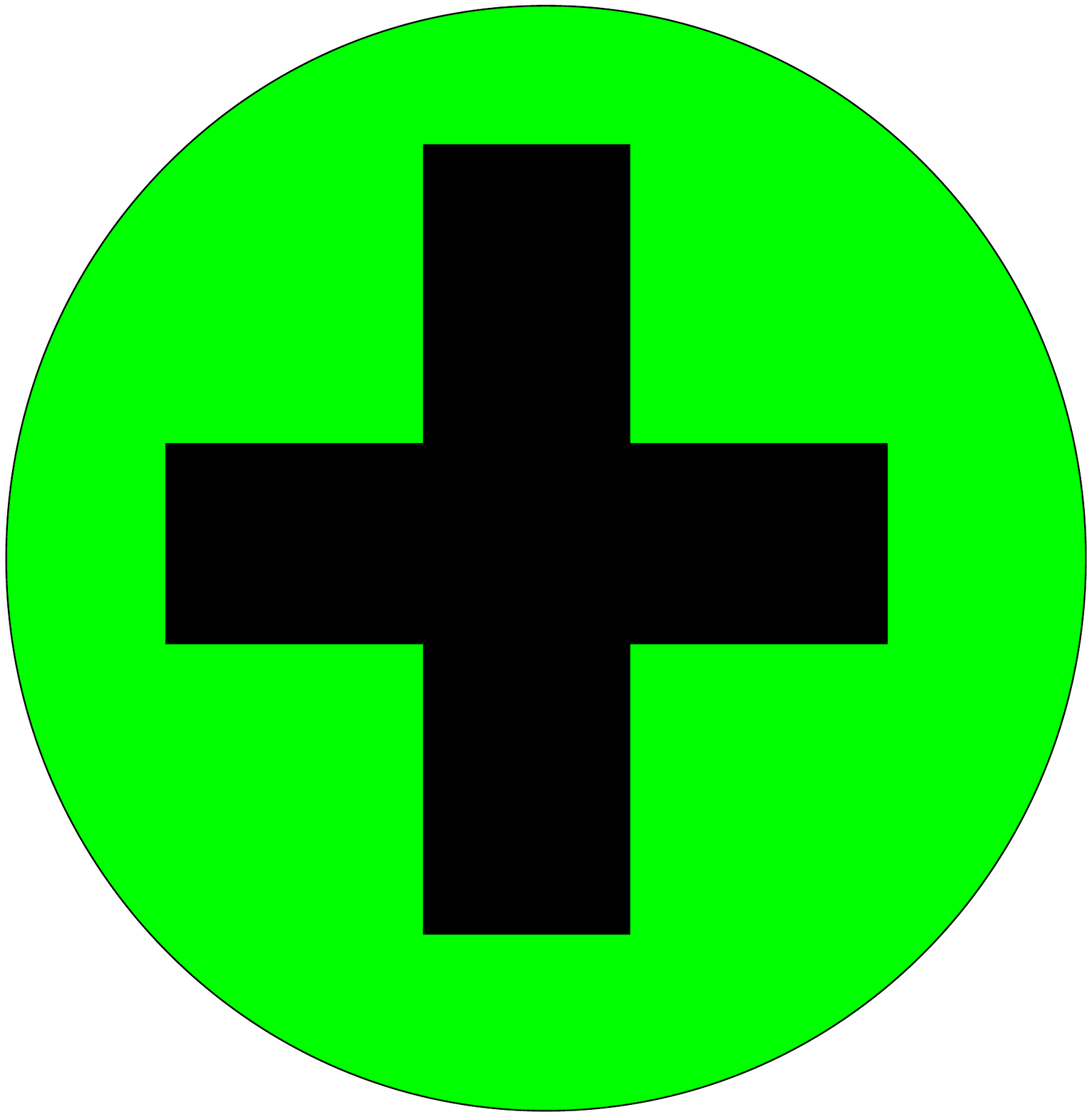}}%
\begin{itemize}}{\end{itemize}}

\newenvironment{itemize_negative}{%
\renewcommand{\labelitemi}{\includegraphics[height=8pt]{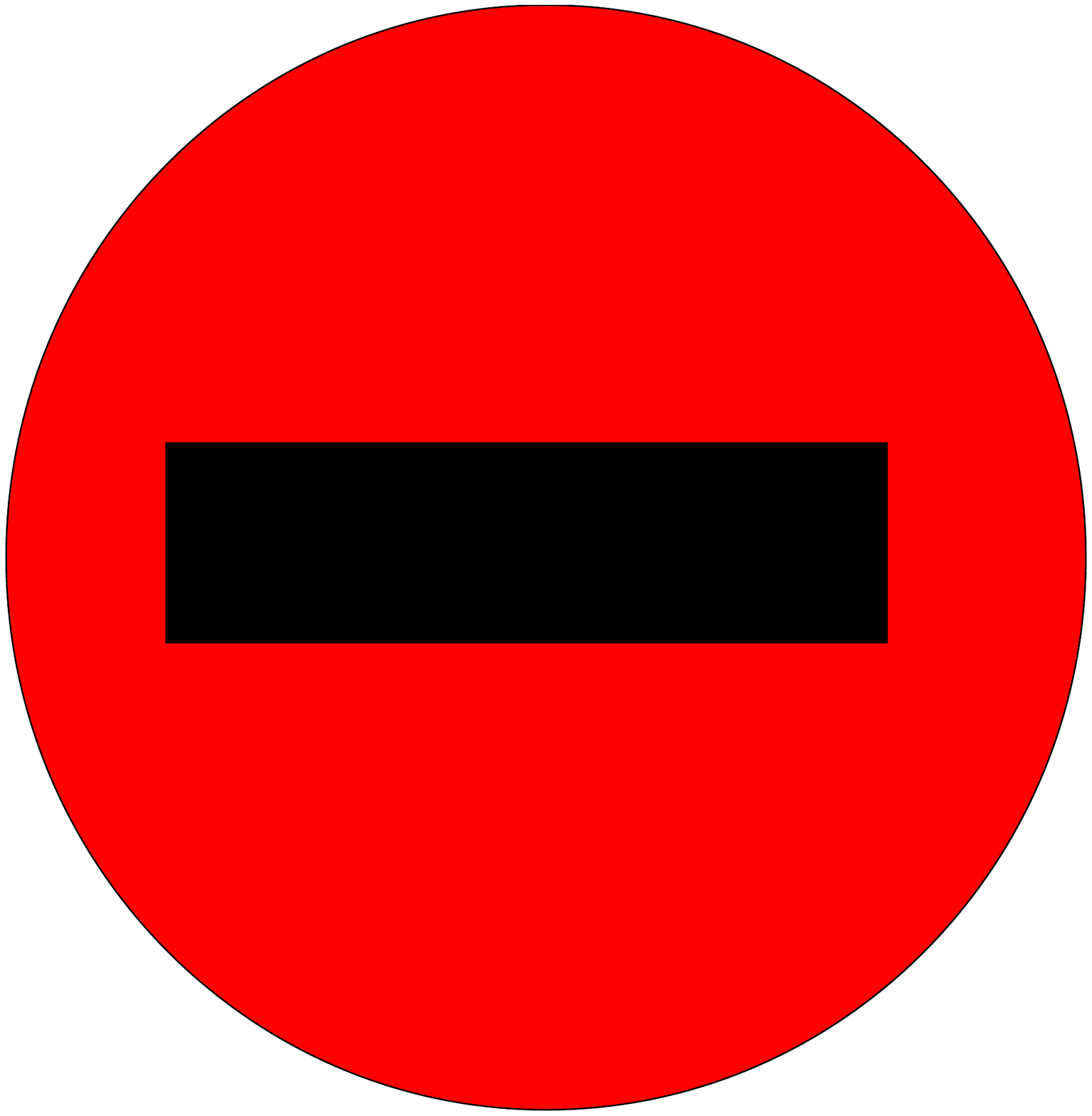}}%
\begin{itemize}}{\end{itemize}}

\title{Rational Unified Process}

\author{\IEEEauthorblockN{Nils~Kopal}
\IEEEauthorblockA{Universität Kassel, Angewandte Informationssicherheit,\\
Pfannkuchstra{\ss}e~1, 34121~Kassel, Germany\\
nils.kopal@uni-kassel.de}}

\maketitle

\begin{abstract}
Bei dieser Seminarausarbeitung handelt es sich um eine Ausarbeitung, die für ein Bachelorkolloquium (Bachelor Angewandte Informatik) an der Universität Duisburg-Essen im Jahr 2011 geschrieben wurde. Ich veröffentliche diese Ausarbeitung, damit interessierte Studierende oder generell an Softwareentwicklung Interessierte sich einen ersten Eindruck über den \emph{Rational Unified Process} erwerben können.

Diese Ausarbeitung bietet eine kurze Einf\"uhrung und einen groben \"Uberblick \"uber den Softwareentwicklungsprozess \textbf{RUP} (\textbf{Rational Unified Process}). Sie beinhaltet einen \"Uberblick \"uber den zugrunde liegenden Prozessaufbau in Kapitel \ref{prozessaufbau}, die Phasen des Prozesses in Kapitel \ref{phasen}, seine Workflows in Kapitel \ref{workflows} und beschreibt die von den RUP Entwicklern immer postulierten ``best practices'' der Softwareentwicklung in Kapitel \ref{best practices}.
\end{abstract}

\section{Einleitung}
In Zeiten immer komplexer werdender Software und immer umfangreicheren Projekten ist es mittlerweile nicht mehr m\"oglich, eine Software nur mit einer kleinen Anzahl an Mitarbeitern und in kurzer Zeit zu entwickeln. In heutigen Softwarefirmen und Projekten ist eine Vielzahl von unterschiedlichem Personal teilweise \"uber Jahre hinweg mit der Erstellung der Software besch\"aftigt. Viele Projekte scheitern noch immer an schlechter Planung, unzureichend oder gar nicht erfassten Anforderungen oder auch an den absolut unterschiedlichen Vorstellungen der Projektbeteiligten. Dies f\"uhrt zu erheblich gesteigerten Entwicklungskosten und vor allem h\"aufig zu frustrierten Mitarbeitern und letztendlich frustrierten Kunden. Um dieser Entwicklung entgegen zu wirken, entwickelt das Software Engineering standardisierte Entwicklungsprozesse. In dieser Ausarbeitung wird versucht, zumindest einen weltweit anerkannten, erprobten und standardisierten Softwareentwicklungsprozess im \"Uberblick zu beschreiben und dem Leser einen Eindruck \"uber diesen zu vermitteln.

\subsection{Einf\"uhrung}

Der \textbf{Rational Unified Process} (kurz \textbf{RUP}) ist ein Softwareentwicklungsprozess, der von der Firma Rational Software entwickelt und vertrieben wurde. Seit 2003 wird der Prozess von IBM, welche Rational Software aufkaufte, unter den alten Produkt- und Firmennamen weiterentwickelt und vertrieben. Erstmals wurde RUP von Daniel Kruchten im Jahr 1996 vorgestellt. Bei RUP handelt es sich nicht um einen einfachen Prozess, sondern vielmehr um ein Prozessframework, das, je nach einsetzender Firma und eingesetztem Projekt, instantiiert und angepasst wird. Im RUP wird die Unified Modeling Language (kurz UML) als Notationsgrundlage genutzt. IBM Rational verkauft RUP als Paket.

\section{Best practices}\label{best practices}
Die Entwickler von RUP sehen diesen als eine Sammlung und Umsetzung von \textbf{``best practices''} (welche ins Deutsche am besten mit ``bew\"ahrten Erfolgsmethoden'' zu \"ubersetzen sind) der Softwareentwicklung. Diese sind hier nun im Folgenden kurz zusammengefasst.

\subsection{Software iterativ entwickeln}

Klassische Softwareentwicklung wurde eher mit sequentiellen Modellen betrieben (wie z.B. das Wasserfallmodell). Hier gibt es eine Reihe von festgelegten Phasen, mit Start- und Endzeitpunkten. Jede Phase besch\"aftigt sich mit einer anderen Thematik (z.B. Analyse, Design, Implementierung, Testen). Problematisch an diesen Modellen ist, dass es meist keinen R\"uckschritt in \"altere Phasen gibt. So m\"ussen Fehler in der Anforderungsanalyse h\"aufig in der Implementierung ``ausgebadet'' werden. Um dies zu vermeiden, wurden iterative Prozesse (wie RUP) entwickelt. Hier durchl\"auft man jede Phase des Prozesses in kleinen Schritten (den Iterationen), um sich so immer n\"aher dem fertigen Produkt zu n\"ahern. Hier ist vorteilhaft, dass auf Fehler oder festgestellte Probleme einfacher reagiert werden kann. Diese werden in einer der nachfolgenden Iterationen korrigiert.

\subsection{Anforderungen verwalten}

Gro\ss e Softwareprojekte scheitern h\"aufig an mangelhaften bzw. schlecht aufgestellten Anforderungen. Teilweise sind den Projektbeteiligten die Anforderungen auch \"uberhaupt nicht oder nur unzureichend bekannt. Um dem entgegenzuwirken, werden Anforderungen an zentraler Stelle verwaltet. So kann sich jeder Projektbeteiligte einen \"Uberblick \"uber selbige machen. \"Anderungen an den Anforderungen werden h\"aufig durch sogenannte \"Anderungsanfragen (engl. Change Request) angefordert und m\"ussen von einem Verantwortlichen zun\"achst gepr\"uft werden. So versucht man, alle Anforderungen konsistent, sinnig und vor allem implementierbar zu halten.

\subsection{Komponentenbasierte Architekturen}

Software in Komponenten zu gliedern, bringt erhebliche Vorteile in der Entwicklung. Durch standardisierte Schnittstellen k\"onnen Komponenten einfach ausgetauscht oder auch wiederverwendet werden. Eine einheitliche Komponentenstruktur macht die Software \"ubersichtlicher, einfacher zu verstehen und somit auch einfacher anzupassen. Das f\"uhrt zu geringeren Entwicklungszeiten und bringt Kostenersparnis mit sich.

\subsection{Visuelle Modellierung}

Visuelle Modellierung vereinfacht die Softwareentwicklung, indem sie Modelle einfacher erstellbar, anpassbar und auch Mitarbeitern ohne technischem Hintergrund verst\"andlicher macht. Mit Hilfe von visueller Modellierung lassen sich Details zun\"achst abstrahieren, um sich auf das Kernproblem zu konzentrieren. Nach und nach k\"onnen diese Modelle dann immer weiter verfeinert werden, bis man fast schon Quellcode erh\"alt (also Quellcode generiert). RUP macht aus diesem Grund erheblich Gebrauch von der UML, da diese standardisierte Modelle f\"ur Use-Cases, Klassen, Komponenten und so weiter mit sich bringt.

\subsection{Software-Qualit\"at sicherstellen}

Qualit\"atssicherung tr\"agt zu einer stabileren, sichereren und korrekteren Software bei, kurz - sie f\"uhrt zu erh\"ohter Qualit\"at und somit zu mehr Kundenzufriedenheit. Mit Hilfe von Komponententests, Integrationstests, Lasttests und kontinuierlichem Abgleich der Software mit ihren Anforderungen und Spezifikation wird versucht, die Qualit\"at auf einem hohem Ma{\ss} zu halten.

\subsection{\"Anderungen an der Software verwalten}

Durch das \"Anderungsmanagement lassen sich \"Anderungen an der zu entwickelnden Software einfach verfolgen. Als Hilfe werden hier unter anderem Versionsverwaltungssysteme, Content-Management-Systeme, Ticketsysteme und dergleichen eingesetzt. Durch Sandboxen erh\"alt jeder Entwickler seinen eigenen, abgeschotteten Bereich in dem er \"Anderungen an der Software vornehmen kann ohne die Entwicklung anderer Projektbeteiligter zu beeinflussen oder gar zu st\"oren. Sind die Entwickler zufrieden mit ihrer Arbeit, k\"onnen sie die \"Anderungen einchecken und so allen anderen verf\"ugbar machen. Im \"Anderungsmanagement werden \"Anderungen h\"aufig durch sogenannte \"Anderungsanfragen (engl. Change requests) begleitet. Diese werden zun\"achst an einem Punkt gesammelt, dann bewertet und angenommen oder abgelehnt. Wurde eine \"Anderungsanfragen angenommen, wird diese einem Entwickler zugeordnet, der die \"Anderung an der Software vornimmt. Im \"Anderungsmanagement werden Richtlinien bez\"uglich der Entwicklungsstruktur und der Nutzung der \"Anderungsmanagementsoftware festgelegt, an die sich jeder Projektbeteiligte h\"alt. So ergibt sich eine durchg\"angige Struktur, die jeder Projektbeteiligte kennt und versteht. So wird die Entwicklung im Team erleichtert und beschleunigt.

\section{Prozessaufbau}\label{prozessaufbau}

\subsection{Aktivit\"aten, Artefakte und Worker}

\begin{figure}[htbp]
\begin{center}
  \includegraphics[width=\columnwidth]{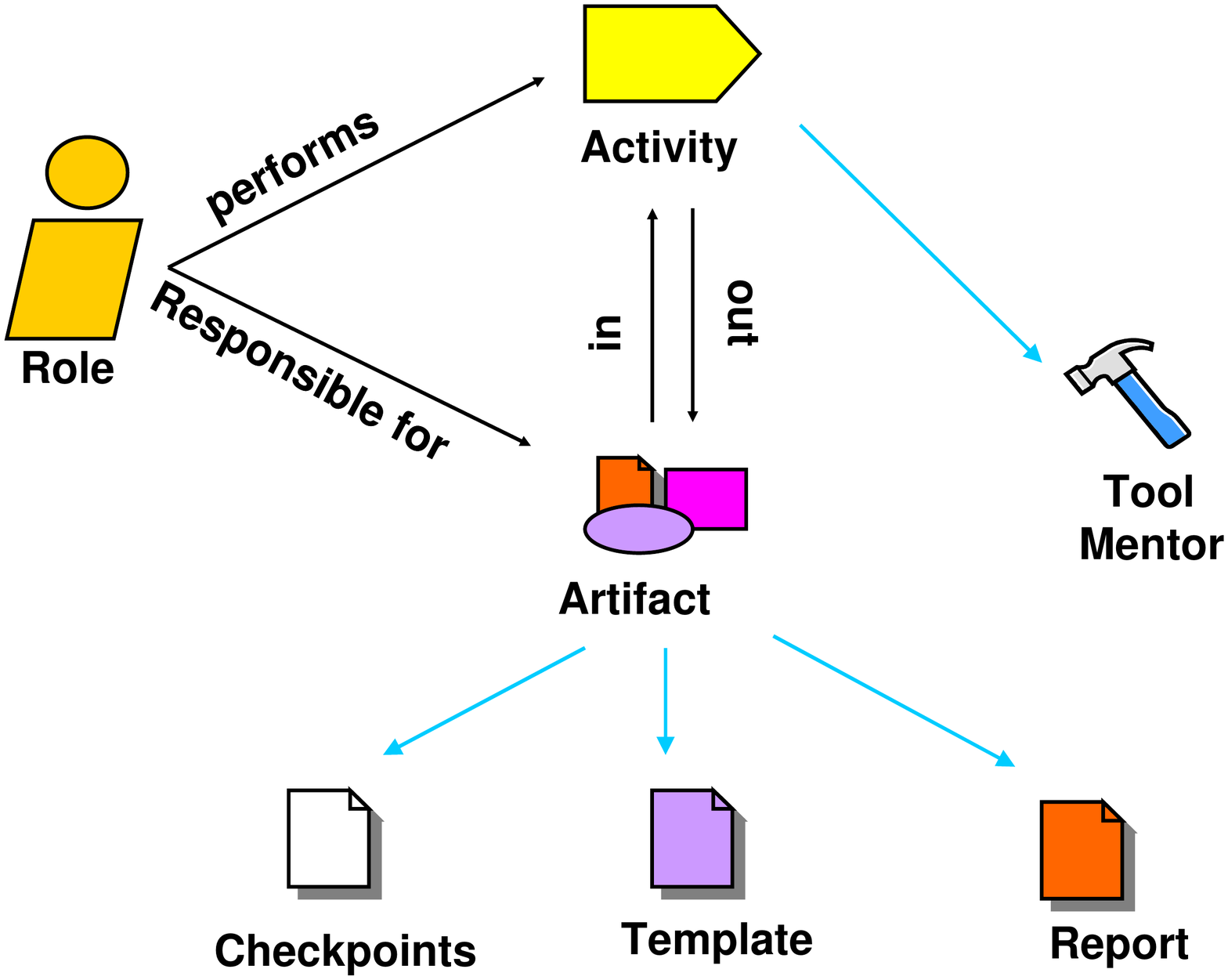}
  \caption[RUP \"Ubersicht]{RUP \"Ubersicht -- entnommen von \url{http://www.ibm.com/developerworks/rational/library/oct07/uttangi_rizwan/index.html}}
  \label{rup-role_activity_artifact}
\end{center}
\end{figure}

Im RUP gibt es eine Reihe von Fachtermini, wobei hier drei der wichtigsten n\"aher beschrieben werden. Zum einen werden alle Ausgabeprodukte, die in RUP produziert werden, als sogenannte \textbf{Artefakte} bezeichnet. Hierbei kann es sich um ein Dokument, Sourcecode, ein Diagramm oder \"ahnliches handeln. Alles was ``anfassbar'' ist, wird als Artefakt bezeichnet.

Desweiteren wird im RUP von \textbf{Workern} gesprochen. Ein Worker ist eine Rolle im RUP. Dies kann z.B. ein Prozess-Designer, ein Softwarearchitekt und so weiter sein.

Zuletzt wird man h\"aufig von \textbf{Aktivit\"aten} h\"oren. Aktivit\"aten sind, wie der Name vermuten l\"asst, T\"atigkeiten, die von einem Worker ausgef\"uhrt werden und auch immer einem oder mehrerer Worker zugeordnet sind. Die Workflows des RUP lassen sich auf einzelne Aktivit\"aten herunterbrechen. In Abbildung~\ref{rup-role_activity_artifact} sind die Zusammenh\"ange noch einmal grafisch verdeutlicht.

\subsection{Struktur von RUP}

\begin{figure}[htbp]
\begin{center}
  \includegraphics[width=\columnwidth]{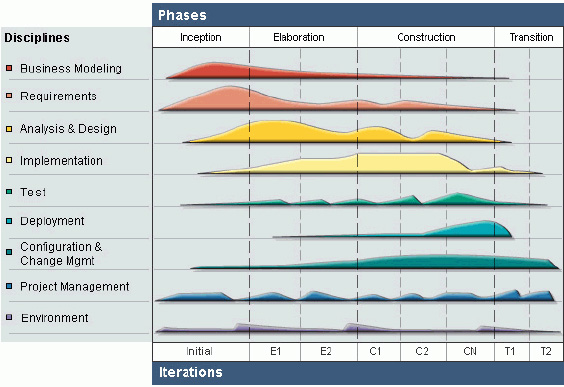}
  \caption[The Rational Unified Process (RUP)]{The Rational Unified Process (RUP) -- entnommen von  \url{http://www.ibm.com/developerworks/webservices/library/ws-soa-term2/}}
  \label{rup-overview}
\end{center}
\end{figure}

RUP ist ein ``zweidimensionaler'' Proze\ss . In der Abbildung \ref{rup-overview} lassen sich auf der horizontalen Achse die vier Phasen des Prozesses erkennen. Auf der vertikalen Achse kann man die neun im RUP definierten Workflows erkennen.

\subsection{Phasen}\label{phasen}

\begin{figure}[htbp]
\begin{center}
  \includegraphics[width=\columnwidth]{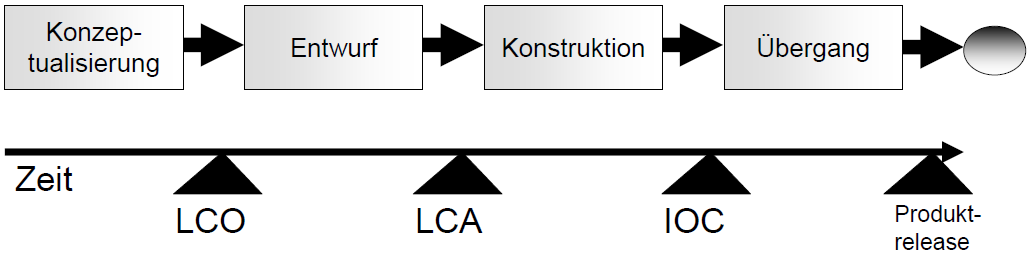}
  \caption[Die vier Phasen und die Meilensteine des Iterativen Prozesses]{Die vier Phasen und die Meilensteine des Iterativen Prozesses -- entnommen von \cite{kruchten}}
  \label{meilensteine}
\end{center}
\end{figure}

Es existieren vier grundlegende \textbf{Phasen} im RUP, die sequentiell durchgearbeitet werden. Diese werden jedoch wiederum in beliebig viele \textbf{Iterationen} unterteilt (je nach Umfang des einsetzenden Projekts). Eine jede Iteration beinhaltet die sogenannten \textbf{Kern-Workflows}, grundlegende T\"atigkeiten, die in jeder Phase durchgef\"uhrt werden. Zeitlich gliedert sich RUP in vier benannte Phasen (\textbf{Konzeptualisierungsphase, Entwurfsphase, Konstruktionsphase, \"Ubergangsphase}) die in diesem Kapitel n\"aher beschrieben werden. Jede Phase wird in einer bis beliebig vielen Iterationen durchlaufen. Dies ist von der Art und dem Umfang des Projektes abh\"angig. In jeder Iteration werden alle Workflows durchgearbeitet. Je nach Phase sind die Anteile der Workflows unterschiedlich gro\ss, wie man es auch in Abbildung \ref{rup-overview} an der unterschiedlichen Dicke der horizontalen Linien je Phase, erkennen kann. Der Abschluss einer jeden Phase bildet das Erreichen eines Milestones (erkennbar in Abbildung \ref{meilensteine}).

\subsubsection{Die Konzeptualisierungsphase}

Die \textbf{Konzeptualisierungsphase} (engl. \textbf{Inception-phase}) dient laut \cite{kruchten} dazu, alle Projektbeteiligten auf ein gemeinsames Projektziel einzustimmen.

Hierzu geh\"oren das Verstehen und Aufstellen der Anforderungen, die an das zu entwickelnde System gestellt werden. Dazu werden f\"ur jede identifizierte Anforderung Anwendungsf\"alle (in Form von Use-Case-Diagrammen mit textueller Beschreibung) aufgestellt.  Ein weiteres Ziel ist das Finden einer Architektur, mit eventueller Entwicklung eines oder mehrerer Prototypen. Am Ende der Phase sollte ein Zeitplan und eine Kostenabsch\"atzung f\"ur die folgenden Phasen stehen. Zu guter Letzt werden in der Konzeptualisierungsphase etwaige Risiken abgesch\"atzt, erkannt und dokumentiert.

Als Ausgabe-Artefakte sieht \cite{kruchten} folgende Dokumente:
\begin{itemize}
 \item Ein \textbf{Visionsdokument}, das die Ideen bez\"uglich der wesentlichen Funktionen des Systems beschreibt
 \item Ein \textbf{\"Uberblicks-Use-Case-Modell}, das alle bereits identifizierten Anwendungsf\"alle und deren Akteure auflistet
 \item Ein \textbf{Projektplan}, der die geplanten Iterationen darstellt
 \item Ein erstes \textbf{Glossar} wird w\"ahrend der Phase angelegt und \"uber alle Phasen hinweg weitergepflegt
\end{itemize}

Am Ende der Konzeptualisierungsphase steht der \textbf{Milestone LCO (Lifecycle Objectives Milestone)}. Hier sollte laut \cite{kruchten} folgendes abgepr\"uft werden:

\begin{itemize}
 \item Das Projektverst\"andnis aller Projektbeteiligten sollte auf einer Linie sein (bez\"uglich Umfangsdefinition, Projektkosten und Einsch\"atzung des Zeitplans)
 \item Die Anforderungen sollten von allen Projektbeteiligten verstanden sein
 \item Glaubw\"urdigkeit von Kosten- und Zeitplaneinsch\"atzungen, Priorit\"aten, Risiken und des Entwicklungsprozesses
 \item Detaillierungsgrad jedes Architekturprototyps, der entwickelt wurde
 \item Die aktuellen Ausgaben im Vergleich zu den vorhergesagten
\end{itemize}

\cite{kruchten} meint, dass das Projekt, falls es an diesem Milestone scheitert entweder abgebrochen oder gr\"undlich \"uberdacht werden sollte.

\subsubsection{Die Entwurfsphase}

Die \textbf{Entwurfsphase} (engl. \textbf{Elaboration-phase}) entwickelt die in der Konzeptualisierungsphase begonnenen Anwendungsf\"alle weiter und bringt diese auf einen m\"oglichst festen Stand (laut \cite{essigkrug} auf eine Vollst\"andigkeit von ca 80\%). Die Architektur sollte am Ende dieser Phase vollst\"andig beschrieben sein und in einem Architekturprototyp vorliegen. Dieser dient als Grundlage f\"ur die fortschreitende Entwicklung der darauffolgenden Phasen.\\
Als Ausgabe-Artefakte sieht \cite{kruchten} folgende:

\begin{itemize}
 \item Den bereits beschrieben \textbf{Architekturprototyp} nebst Beschreibung der Architektur
 \item Ein zu 80\% fertiges \textbf{Use-Case-Modell}
 \item \"Uberarbeitete \textbf{Projekt- und Entwicklungspl\"ane }und \textbf{Risikenabsch\"atzungen}
\end{itemize}

Am Ende der Entwurfsphase steht der zweite Milestone \textbf{LCA (Lifecycle Architecture)}. Hier werden laut \cite{kruchten} folgende Dinge abgepr\"uft:

\begin{itemize}
 \item Stabilit\"at der Vision des zu entwickelnden Produkts
 \item Stabilit\"at der entworfenen Architektur
 \item Wurden alle wesentlichen Risiken gefunden?
 \item Detailgrad der Planung f\"ur die Konstruktionsphase ausreichend?
 \item Erreichbarkeit der Vision mit Hilfe der entwickelten Architektur und Pl\"ane
 \item Akzeptanz der aktuellen Kosten im Vergleich zu geplanten Kosten
\end{itemize}

\cite{kruchten} meint, dass das Projekt, falls es an diesem Milestone scheitert entweder, wie im ersten Milestone, abgebrochen oder \"uberdacht werden sollte.

\subsubsection{Die Konstruktionsphase}
Die \textbf{Konstruktionsphase} (engl. \textbf{Construction-phase}) dient der vollst\"andigen Entwicklung und dem Testen der Komponenten des Systems. Ziel der Phase ist eine fertige und einsetzbare Sofware.\\

Als Ausgabe-Artefakte sieht \cite{kruchten} mindestens:

\begin{itemize}
 \item Das fertig entwickelte \textbf{System}
 \item Eine \textbf{Releasebeschreibung} des aktuellen Releases
 \item Ein \textbf{Benutzerhandbuch}
\end{itemize}

Am ende der Konstruktionsphase steht der Milestone \textbf{IOC (Initial Operational Capability)}. Bei diesem dritten Milestone wird folgendes abgepr\"uft:

\begin{itemize}
 \item Ist die Produktstabilit\"at ausreichend damit die Software beim Kunden eingesetzt werden kann?
 \item Sind alle Projektbeteiligten soweit in den Betrieb beim Kunden \"uberzugehen?
 \item Akzeptanz der aktuellen Kosten im Vergleich zu geplanten Kosten
\end{itemize}

Genau wie in den vorherigen Milestones f\"uhrt eine Nichterreichung des Milestones zu einem Projektabbruch oder zu einer gr\"undlichen \"Uberdenkung des Projektes.

\subsubsection{Die \"Ubergangsphase}

Ziel der \textbf{\"Ubergangsphase} (engl. \textbf{Transition-phase}) ist die Integration des fertigen Systems in die Umgebung des Kunden. Hierzu werden Beta-Tests angesetzt. Das System wird an die bestehenden Systeme des Kunden angeschlossen (Datenbanksysteme, Netzwerk etc.) - also installiert. Gleichzeitig werden Benutzer und Administratoren auf Seiten des Kunden mit dem Umgang des Systems geschult. Etwaige Fehler werden korrigiert und in neu erstellten Releases behoben. Am Ende dieser Phase soll es dem Kunden m\"oglich sein, das System ohne weitere Hilfe des Entwicklers selbst betreiben zu k\"onnen. Zusammen mit dem Kunden wird festgehalten, ob und wie das System die an es gestellten Anforderungen erf\"ullt.

Der letzte Milestone \textbf{Produktrelease} steht am Ende dieser Phase. Hier wird noch folgendes abgepr\"uft:

\begin{itemize}
 \item Ist der Kunde mit dem Ergebnis des Projektes zufrieden?
 \item Akzeptanz der aktuellen Kosten im Vergleich zu geplanten Kosten
\end{itemize}

\subsection{Workflows}\label{workflows}

Die \textbf{Workflows} sind grundlegende T\"atigkeiten, die in jeder Phase (und in jeder Iteration) durchgef\"uhrt werden und aus unterschiedlichen Aktivit\"aten bestehen. Jedem Workflow sind bestimmte Worker zugeordnet. Wie bereits beschrieben, werden die einzelnen Workflows, je nach Phase, in unterschiedlich starker Auspr\"agung ausgef\"uhrt. Im RUP gibt es neun sogenannte Kern Workflows, die wiederum in zwei unterschiedliche Arten eingeteilt sind:

\begin{itemize}
 \item Die \textbf{Engineering Workflows}, von denen es sechs gibt
 \item Die \textbf{Supporting Workflows}, von denen es drei gibt
\end{itemize}

Ein Workflow kann ein oder mehrere Artefakte als Ausgabe besitzen (z.B. ein Use-Case-Modell, das die Anforderungen beschreibt beim Anforderungs-Workflow). Die Workflows sind in Abbildung \ref{rup-overview} als vertikale Achse zu erkennen.

\subsubsection{Engineering Workflows}
Die \textbf{Engineering Workflows} sind, wie der Name vermuten l\"asst, die ``technischen'' Workflows im RUP. Sie besch\"aftigen sich mit der Entwicklung und der Unterst\"utzung der Entwicklung des Systems.

\begin{figure}[htbp]
\begin{center}
  \includegraphics[width=0.75\columnwidth]{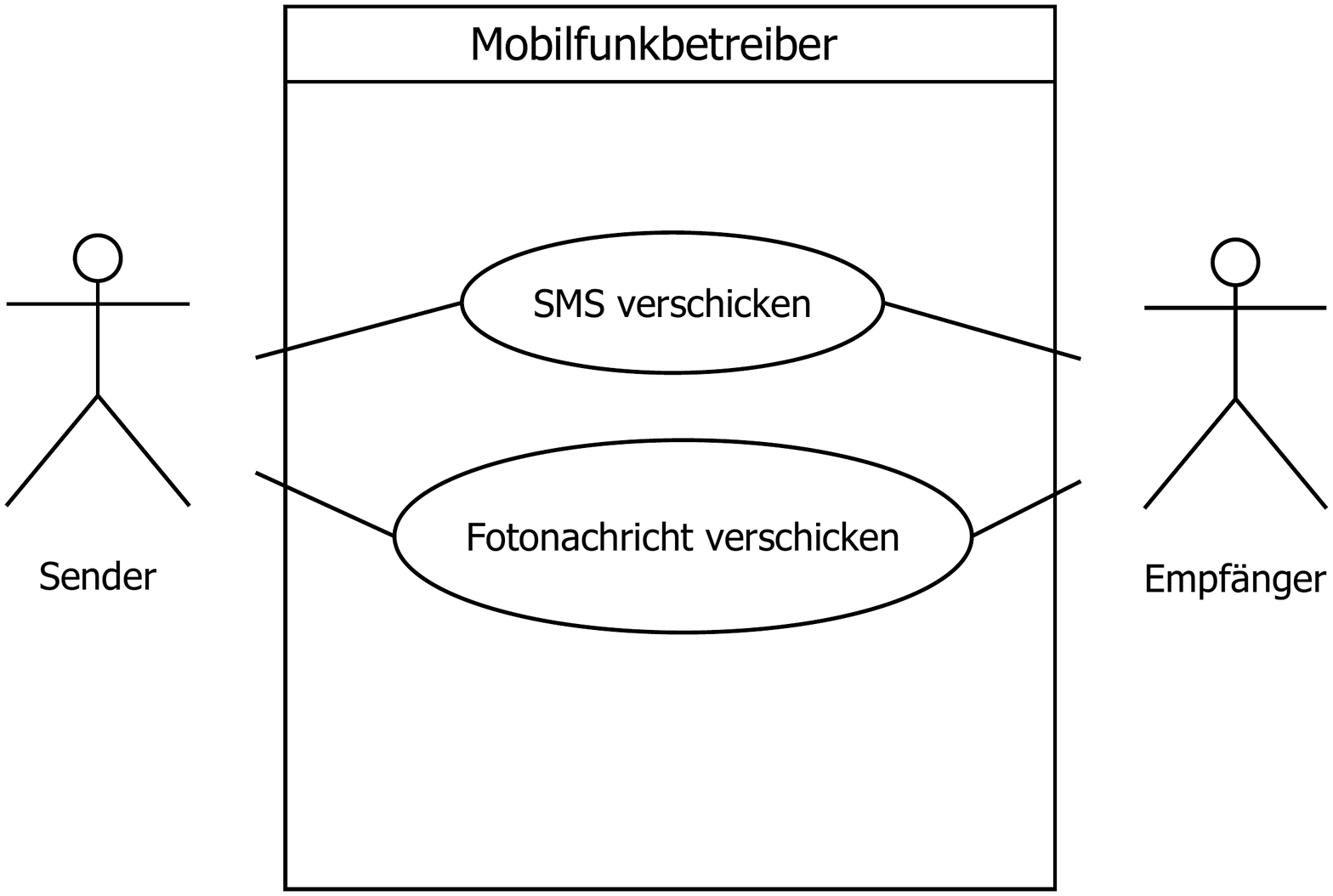}
\end{center}
\caption{Beispiel eines Use-Cases}
\label{anwendungsfall}
\end{figure}

\paragraph{Gesch\"aftsprozessmodellierungs-Workflow}

Im \textbf{Gesch\"aftsprozessmodellierungs-Workflow} (engl. \textbf{Business modeling workflow}) wird ein Modell der Organisation und der Prozesse des Kunden entwickelt. Dies wird in einem \textbf{Gesch\"afts-Use-Case-Modell} (ein Beispiel f\"ur einen Anwendungsfall sieht man in Abbildung \ref{anwendungsfall} ) und einem Objektmodell festgehalten. \cite{kruchten} meint, dass dies nicht in jedem Softwareprojekt notwendig sei. Als Beispiel nennt er hier die \"Anderung der Funktionalit\"at einer Telekommunikationsanlage, da hier keine vorherige Gesch\"aftsprozessmodellierung notwendig sei. Ein Beispiel, wo dies notwendig sei, w\"are z.B. die Entwicklung einer Auftragsverwaltung f\"ur die Vertriebsmitarbeiter von Telekommunikationsanlagen. Der Vertriebs- und Bestellprozess sei in diesem Kontext sehr komplex, da es sich oft um kundenspezifische L\"osungen und nicht um Fertigprodukte handele. Hier macht eine Gesch\"aftsprozessmodellierung auf jeden Fall Sinn.

An diesem Workflow sind im Wesentlichen zwei Arten von Worker beteiligt:

\begin{itemize}
 \item Der \textbf{Gesch\"aftsprozess-Analytiker}, der die Organisationseinheiten des Kunden grob beschreibt und die Gesch\"afts-Use-Cases, und deren Verbindungen zueinander, identifiziert
 \item Der \textbf{Gesch\"aftsprozessdesigner}, der die Workflows innerhalb der Gesch\"afts-Use-Cases beschreibt und diesen Arbeiter zuordnet.
\end{itemize}

\paragraph{Anforderungs-Workflow}

Der \textbf{Anfoderungs-Worfklow} (engl. \textbf{Requirements workflow}) dient dazu, die Anforderungen an das System aufzustellen.

\cite{kruchten} beschreibt eine Anforderung, als eine Bedingung oder eine Funktionalit\"at, der das System entsprechen muss. Im Anforderungs-Workflow wird versucht, die Anforderungen, die der Kunde an das System hat, aufzustellen. Im RUP werden diese Anforderungen in einem \textit{Use-Case Modell} festgehalten. Weitere Ziele sind, laut \cite{kruchten}, die Beschreibung einer Benutzerschnittstelle f\"ur das System und eine Basis zu schaffen, um die weiteren Iterationen zu planen und um die ben\"otigte Zeit und Kosten abzusch\"atzen.

Als beteiligte Worker nennt \cite{kruchten}:

\begin{itemize}
 \item Den \textbf{Systemanalytiker} als Leiter der Anforderungsermittlung
 \item Den \textbf{Use-Case-Spezifizierer}, welcher die einzelnen Use-Cases genauer definiert
 \item Den \textbf{Architekten}, der die f\"ur die Softwarearchitektur wichtigen Use-Cases identifiziert
\end{itemize}

\paragraph{Analyse- und Design-Workflow}

Im \textbf{Analyse- und Design-Workflow} (engl. \textbf{Analysis- and Design-workflow}) werden die durch den Anforderungs-Workflow gewonnen Anforderungen in Spezifikationen umgewandelt. Das wichtigste Artefakt, das im Analyse- und Design-Workflow produziert wird, ist das Designmodell. Es besteht laut \cite{kruchten} aus Kollaborationen von Klassen, die in Paketen und Subsystemen zusammengefasst werden. Der Analyse- und Design-Workflow bildet den \"Ubergang von den Anforderungen zu der Implementierung. Ziel ist es, ein komponentenbasiertes Design zu entwickeln. Komponenten besitzen definierte Schnittstellen und h\"angen nur von den Schnittstellen andere Komponenten ab. Durch die Nutzung von Komponenten lassen sich die Teile des Systems einzeln entwickeln und unabh\"angig von anderen Teilen austauschen und auch in anderen Systemen wiederverwenden.

Beteiligte Worker sind:

\begin{itemize}
 \item Der \textbf{Architekt}, welcher die Gruppierung aller Elemente (Klassen) und die Definition der Schnittstellen zwischen diesen vornimmt
 \item Der \textbf{Designer} ist f\"ur den Entwurf von Klassen und deren Beziehungen und/oder ganzen Paketen von Klassen verantwortlich
  \item Ist eine Datenbank erforderlich, wird diese vom \textbf{Datenbank-Designer} entworfen.
  \item F\"ur das Review von Architektur und Datenbankdesign sieht RUP \textbf{Architektur- und Datenbankgutachter} vor
\end{itemize}

\paragraph{Implementierungs-Workflow}

Der \textbf{Implementierungs-Workflow} (engl. \textbf{Implementation workflow}) befasst sich mit der konkreten Implementierung der Software, also dem Erstellen von Sourcecode. Dieser wird in Klassen und Komponenten zusammengestellt. Des Weiteren wird der Sourcecode einer jeden Komponente mittels Integration zusammengebracht. Parallel zu der Implementierung laufen in diesem Workflow Komponententests der einzelnen Komponenten ab. Ausgabeartefakte w\"aren zum einen die einzelnen Komponenten als auch die zusammengefassten Komponenten in einem sogenannten Implementierungssubsystem.

Beteiligte Worker an diesem Workflow sind:
\begin{itemize}
  \item Der \textbf{Implementierer}, der die einzelnen Komponenten implementiert
  \item Der \textbf{Integrierer}, der alle Komponenten miteinander verbindet
\end{itemize}

\paragraph{Test-Workflow}

Der \textbf{Test-Workflow} dient im RUP der Qualit\"atssicherung. Im Test-Workflow wird laut \cite{essigkrug} versucht, Fehler aufzudecken und nicht, solche zu beheben oder zu l\"osen. Getestet wird nach den im Anforderungs-Workflow erstellten Anforderungen. Hierf\"ur wird ein Testmodell aufgestellt, das alle Testf\"alle, Testprozeduren, Testscripte und erwarteten Ergebnisse sowie Beschreibungen der Tests beinhaltet. Ausgabeartefakte des Test-Workflows sind ein Testplan, das bereits genannte Testmodell, ein Auslastungsmodell, welches zum Performance-Testen genutzt wird und nat\"urlich Fehler ($=$ Fehlgeschlagene Tests), die in \"Anderungsanfragen (zum Beheben der Fehler) umgewandelt werden.

Beteiligte Worker an diesem Workflow sind:
\begin{itemize}
  \item Der \textbf{Test-Designer}, welcher f\"ur die Planung und Auswertung der Tests zust\"andig ist
  \item Der \textbf{System-Tester}, der die Durchf\"uhrung der Tests vornimmt
  \item Der \textbf{Integrations-Tester}, der alle Komponenten nach der Integration im Zusammenspiel testet
  \item Der \textbf{Performance-Tester}, der, wie sein Name vermuten l\"asst, f\"ur das Testen unter Performance-Gesichtspunkten verantwortlich ist
\end{itemize}

\paragraph{Verteilungs-Workflow}

Der \textbf{Verteilungs-Workflow} (engl. \textbf{Deployment-workflow}) beinhaltet, wie der Name sagt, die Verteilung des fertigen Produkts. RUP unterscheidet laut \cite{essigkrug} zwischen drei Arten der Verteilung: Zum einen die Verteilung als Installation beim Kunden, dann die Verteilung als Standardprodukt und die Verteilung speziell \"uber das Internet. Mit der Verteilung geht das Projekt in den operativen Betrieb \"uber.

Zu den beteiligten Workern z\"ahlt \cite{essigkrug} unter anderem
 
\begin{itemize}
  \item Den \textbf{Deployment Manager}, der unter anderem einen Deployment Plan entwickelt, Release Notes schreibt, das Produkt letztendlich frei gibt sowie Beta- und Akzeptanztests \"uberwacht
  \item Der \textbf{Implementierer}, der die Installationsroutinen oder -Scripte f\"ur die Installation der Software erstellt
\end{itemize}

\subsubsection{Supporting-Workflows}

Neben den technischen Workflows gibt es im RUP noch drei \textbf{Supporting-workflows}. Diese ``produzieren'' nicht, wie die Engineering Workflows, sondern dienen als Unterst\"utzende Workflows allen Projektbeteiligten bei ihrer Arbeit.

\paragraph{Projektmanagement-Workflow}

Der \textbf{Projektmanagement-Workflow} dient, wie auch das Projektmanagement im Allgemeinen in anderen Projekten, der Steuerung und Kontrolle des Projektflusses. Hierzu z\"ahlen die Planung aller am Projekt beteiligten Ressourcen, der Zeit sowie Kontrolle der Einhaltung des Projektplans. Laut \cite{essigkrug} ist hier das Risikomanagement von besonderer Bedeutung. Hierunter fallen die Aufstellung und Bewertung einer Liste von Projektrisiken, um diesen fr\"uh genug entgegenwirken zu k\"onnen.

Der Wichtigste am Projektmanagement Workflow beteiligte Worker ist der \textbf{Projektmanager}, der planen, kontrollieren und gegebenenfalls (fr\"uh genug) in das Projekt eingreifen muss, wenn es zu Problemen kommt. Desweiteren verteilt er die Aufgaben im Projekt, legt den Grundstein f\"ur das Projekt mit dem Projektstart, berichtet an die Auftraggeber des Projekts (z.B. die Gesch\"aftsleitung) \"uber den aktuellen Status, hat einen \"Uberblick \"uber die Projektkosten, kontrolliert alle Projektbeteiligten routinem\"assig und beauftragt Projektmitglieder mit Reviews.

\paragraph{Konfigurations- und \"Anderungsmanagement-Workflow}
Im \textbf{Konfigurations- und \"Anderungsmanagement-Workflow} (engl. \textbf{Configuration- and Changemanagement-workflow})wird versucht, alle im Projekt produzierten Artefakte und ihre Beziehungen zueinander zu sichern. Sichern bedeutet hier, sowohl ihren aktuellen Zustand als auch alle vorherigen Zust\"ande und deren \"Anderungen an einem zentralen Punkt zu speichern. So ist es allen Projektbeteiligten m\"oglich, \"Anderungen sowohl selbst einfach zu t\"atigen und an alle anderen zu propagieren, als auch \"Anderungen anderer einfach nachzuvollziehen.

Als wichtige neue Worker findet man in diesem Workflow unter anderem
\begin{itemize}
  \item Den \textbf{Konfigurationsmanager}, der die Projektstruktur innerhalb des Konfigurationsmanagementsystems festlegt und dem Projektmanager \"uber selbiges berichtet
  \item Alle anderen Projektmitglieder, da sie das Konfigurationsmanagementsystem als Werkzeug nutzen
\end{itemize}

\paragraph{Umgebungs-Workflow}
Der \textbf{Umgebungs-Workflow} (engl. \textbf{Environment workflow}) dient dem Bereitstellen einer Entwicklungsumgebung und Prozessen, um die Entwicklung zu unterst\"utzen. Es werden Standards festgelegt, z.B. welche Entwicklungsumgebung und Tools genutzt werden. Au\ss erdem dient der Workflow dem konkreten Anpassen von RUP selbst f\"ur das aktuell laufende Projekt. Als wichtige Workflowbeteiligte lassen sich \textbf{Tool-Spezialisten} nennen. Diese m\"ussen Tools ausw\"ahlen und auf ihre Eignung f\"ur das Projekt pr\"ufen.

\section{Fazit}

Dieses Kapitel gibt einen stichwortartigen \"Uberblick \"uber Vorteile und Nachteile, die der Einsatz von RUP mit sich bringt. Gegen Ende steht eine kurze Zusammenfassung dieser Ausarbeitung.

\subsection{Vorteile}
\begin{itemize_positive}
  \item \textbf{Bekanntheitsgrad}: RUP ist ein weltweit verbreiteter Prozess, der schon \"uber Jahre hinweg eingesetzt und weiterentwickelt wird
  \item \textbf{Literatur}: Es gibt unz\"ahlige Literatur, die sich mit RUP besch\"aftigt
  \item \textbf{Toolunterst\"utzung}: Der Prozess wird ``von Haus aus'' mit Tools unterst\"utzt (die von IBM Rational entwickelt werden)
  \item \textbf{Anpassbarkeit}: RUP ist individuell an die Projektstruktur und Organisationsstruktur anpassbar
  \item \textbf{Iterativer Prozess}: Durch iteratives Vorgehen kann sich ``Schritt f\"ur Schritt'' an die Probleml\"osung getastet werden. Probleme k\"onnen fr\"uhzeitig erkannt und behoben oder von vornherein vermieden werden
  \item \textbf{Unified Modeling Language}: RUP setzt UML als Notation ein - somit sind Teile von RUP bereits meist ohne Kenntnis des eigentlichen Prozesses bekannt. Eine Kooperation mit externen Schnittstellen, wie anderen Unternehmen oder dem Kunden, wird so vereinfacht
  \item \textbf{Weiterentwicklung}: IBM Rational entwickelt kontinuierlich am Prozess und versucht diesen so mit aktuellen Neuentwicklungen und dem neuesten Stand der Softwareentwicklung auszur\"usten
  \item \textbf{Skalierbarkeit}: RUP ist sowohl f\"ur kleine Projekte als auch f\"ur gro\ss e Projekte flexibel einsetzbar. Der Nutzer w\"ahlt vor der Instanziierung des Prozesses die ihm sinnvoll erscheinenden Teile und nutzt diese im Projektverlauf
  \item \textbf{Klare Struktur}: RUP bietet eine \"ubersichtliche, klare Struktur, was den Aufbau des Prozesses als auch die Verantwortlichkeiten w\"ahrend der Ausf\"uhrung des Prozesses betrifft
\end{itemize_positive}

\subsection{Nachteile}
\begin{itemize_negative}
  \item \textbf{Nicht kostenlos}: RUP ist propriet\"ar und kostet Geld
  \item \textbf{Schwierige Einf\"uhrung}: F\"ur die erfolgreiche Einf\"uhrung von RUP sollte man externe Berater zwingend als Unterst\"utzung einsetzen
  \item \textbf{Komplexit\"at}: RUP ist ein \"uberaus m\"achtiges Werkzeug, dessen Handhabung Jahre der Erfahrung und Nutzung voraussetzt
  \item \textbf{RUP muss gelebt werden}: Wie jeder andere Softwareentwicklungsprozess muss RUP von jedem Projektbeteiligten verstanden, akzeptiert und durchgef\"uhrt werden
\end{itemize_negative}

\subsection{Zusammenfassung}
RUP ist ein iterativer, inkrementeller, UML-gest\"utzter Softwareentwicklungsprozess. Er ist sogar mehr als das, er ist ein Prozessframework, dass je nach nutzendem Projekt individuell instanziiert werden muss. Grundlegend ist RUP in vier Phasen aufgebaut (Konzeptualisierungsphase, Entwurfsphase, Konstruktionsphase, \"Ubergangsphase). Am Ende einer jeden Phase steht ein Milestone der erreicht werden muss.

RUP definiert Worker, Artefakte und Aktivit\"aten. Ein Worker ist eine Rolle, die einer oder mehreren Aktivit\"aten zugeordnet ist. Aktivit\"aten sind grundlegende T\"atigkeiten, wie z.B. Programmieren, Dokumentieren und so weiter. Das Ergebnis einer Aktivit\"at sind meistens ein oder mehrere Artefakte. Als Artefakt wird all jenes, das ``anfassbar'' ist bezeichnet. Beispiele hierf\"ur w\"aren ein Dokument, Sourcecode, Projektpl\"ane und so weiter.

Jede Phase von RUP wird in beliebig viele Iterationen aufgeteilt. Die Anzahl der Iterationen je Phase h\"angt vom Umfang des Projektes ab f\"ur das RUP eingesetzt wird. W\"ahrend einer jeder Iteration werden die neun Kern Workflows in unterschiedlichen Anteilen durchlaufen.

Ein Workflow ist eine Sammlung von Aktivit\"aten und wird einem oder mehreren Workern zugeordnet. Es gibt sechs Engineering Workflows, die technisch getrieben sind, und drei Supporting Workflows, die den Prozess insgesamt unterst\"utzen.

Dank seiner Individualisierbarkeit ist RUP sowohl f\"ur kleine und mittlere als auch gro\ss e Softwareprojekte geeignet.

Die sechs ``best practices'' der Softwareentwicklung dienten bei der Entwicklung von RUP als Grundlage.

\nocite{*}
\bibliographystyle{IEEEtran}
\bibliography{IEEEabrv,references}

\end{document}